\documentclass[aps,twocolumn,preprintnumbers,amsmath,amssymb,prb,superscriptaddress,longbibliography]{revtex4-2}

\usepackage{amssymb}

\usepackage{color}
\usepackage{amsmath}
\usepackage{graphicx}
\usepackage[breaklinks,colorlinks,bookmarks=false,citecolor=blue,linkcolor=blue,urlcolor=blue]{hyperref}
\usepackage{braket}
\usepackage{float}
\usepackage{fancyhdr}
\usepackage{subcaption}

\usepackage{natbib,hyperref} 
\usepackage{soul} 

\begin{document}%

\preprint{\href{https://doi.org/10.1103/PhysRevB.110.045150}{Phys. Rev. B {\bfseries 110}, 045150 (2024)}}

\newcommand{\mycomment}[1]{}

\title{Electronic structure and conductivity in functionalized multilayer black phosphorene}

\author{Jouda Jemaa \surname{Khabthani}}
\email{jouda.khabthani@fst.utm.tn}
\affiliation{
Laboratoire de Physique de la Mati\`ere Condens\'ee,
D\'epartement de Physique, Facult\'e des Sciences de Tunis, Universit\'e El Manar, Campus universitaire,  Tunis, 1060, Tunisia}

\author{Khouloud \surname{Chika}}
\email{khouloud.chika@fst.utm.tn}
\affiliation{
Laboratoire de Physique de la Mati\`ere Condens\'ee,
D\'epartement de Physique, Facult\'e des Sciences de Tunis, Universit\'e El Manar, Campus universitaire,  Tunis, 1060, Tunisia}

\author{Ghassen \surname{Jema\"{i}}}
\email{ghassen.jemai@fst.utm.tn}
\affiliation{
Laboratoire de Physique de la Mati\`ere Condens\'ee,
D\'epartement de Physique, Facult\'e des Sciences de Tunis, Universit\'e El Manar, Campus universitaire,  Tunis, 1060, Tunisia}

\author{Didier \surname{Mayou}}
\email{didier.mayou@neel.cnrs.fr}
\affiliation{
Institut NEEL, CNRS, Univ. Grenoble Alpes, Institut NEEL, 38042 Grenoble, France}

\author{Guy \surname{Trambly de Laissardi\`ere}}
\email{guy.trambly@cyu.fr}
\affiliation{ 
Laboratoire de Physique Th\'eorique et Mod\'elisation,
CY Cergy Paris Universit\'e, CNRS, 
95302 Cergy-Pontoise, France}

\date{\today}

\begin{abstract}

Phosphorene and its components are highly reactive to oxygen when exposed to ambient conditions due to the presence of lone pairs of electrons on phosphorus atoms. Functionalization serves as a solution to prevent the chemical degradation of these materials. In this paper, we investigate the impact of relatively strong covalent or noncovalent functionalization on phosphorene (monolayer black phosphorus (BP)), few-layer BP, and bulk BP. We use an effective tight-binding Hamiltonian that corresponds to one orbital per site, wherein covalent functionalization is simulated by atomic vacancies, and noncovalent functionalization is simulated by Anderson disorder. We demonstrate that these two types of functionalization act differently on the electronic structure 
and quantum diffusion, particularly affecting the gap and mobility characteristics, especially with a high degree of functionalization.
However, we also show that the mobility gap is not significantly modified by the two types of defect.
We also analyze the electron-hole asymmetry that is more important for multilayer and bulk BP.
\end{abstract}

\maketitle

\section{Introduction}\label{sec1}
The scientific community has shown keen interest in two-dimensional (2D) materials, focusing on their potential applications in electronics, sensing, dye degradation, and various biomedical uses \cite{CORTESARRIAGADA2022,KUMBHAKAR2023,MARIAN2022,Long2018}. 
An initial 2D crystal realized by mechanical exfoliation was graphene. 
It is an interesting material for its outstanding electronic and mechanical properties. However, its zero band gap limited its use in optoelectronics and semiconductor devices. Therefore, several research studies have been carried out to find other 2D materials with a band gap and large carrier mobility to replace graphene. In 2014, a successful exfoliation of few-layer black phosphorus (BP), the most stable allotrope of phosphorus, was realized, and phosphorene, i.e., monolayer black phosphorus, was obtained \cite{Li2014,Liu2014,Xia2014}. In contrast to graphene, phosphorene is a 2D semiconductor with a direct band gap with a high hole mobility up to 1000 cm$^2$V$^{-1}$s$^{-1}$ at room temperature \cite{sorkin2017recent}. Its band gap, \(E_g\), varies from 2 eV \cite{Tran2014} for phosphorene monolayer to 0.3 eV for the bulk \cite{Qiao2014,Gaufres19}. 
Monolayer and multilayer BPs exhibit a unique in-plane structural anisotropy which reflects in the band anisotropy around the gap and hence in the effective masses of electrons and holes\,\cite{Qiao2014,Carvalho2016,Xia2014,Akhtar2017}.

Functionalization or disorder can modify the band gap value to tailor these materials for desired applications. Phosphorene and its components are highly sensitive to the environment. This sensitivity arises from their pronounced reactivity to oxygen under ambient conditions, which may constrain their practical use in real devices. Therefore, it is crucial to investigate the impact of defects and impurities on the electronic structure and transport properties for this purpose \cite{DENIS2022}.

Several recent experiments have been carried out to protect few-layer BP from chemical degradation due to the environment and tailor its electrical properties. Covalent and noncovalent functionalization have been realized experimentally to enhance its environmental stability \cite{Mitrovic21,ryder2016,tang2017fluorinated,jellett2020prospects,van2018covalent,yan2019hydroxyl}. The lone pair electrons of phosphorus atoms create a chemical bond with the adatoms. Additionally, theoretical investigations have contributed to the development of the chemistry of few-layer BP using DFT models \cite{Bui18,kuklin2019spontaneous,sun2019transport}. Yet, crucial aspects of its reactivity remain largely unexplored, and there is a lack of consensus between different interpretations of the experimental results. Functionalization can be carried out on the surface of few-layer BP or that of phosphorene, but can also be randomly distributed among the layers \cite{liu2021chemical,antonatos2022black}. It has also been demonstrated that functionalization can be covalent, with resonant scatterers, such as functional groups containing free radicals like hydrogen, halogens, hydroxyl, thiol groups, and organic substituents, which are important for biological applications \cite{taylor2020interplay, DENIS2022}. Functionalization can also be noncovalent, with nonresonant scatterers such as cesium, potassium, or some organic molecules based on van der Waals interaction \cite{liu2021chemical,bolognesi2019,Chen_2020,wang2017charge}.

In this paper, we focus our study on the electronic structure and the conductivity of functionalized monolayer, few-layer BP, and bulk by using the tight-binding (TB) model developed by Rudenko \textit{et al.} \cite{Rudenko2014,Rudenko2015}. In the case of graphene, the description of its electronic properties at low energy is very simple because it is performed by the TB Hamiltonian involving only one hopping parameter. However, for phosphorene, using this method is very complicated because it has four orbitals per atom, and we must take into account the mixture of states of different symmetries. 
With the Rudenko \textit{et al.} simplified Hamiltonian, we model covalent functionalization using vacancies and noncovalent functionalization through Anderson disorder. 
We discuss the impact of these types of adsorbates on the density of states (DOS), conductivity, and mobility in phosphorene, few-layer BP, and the bulk. Our idea is to compare the effect depending on the number of phosphorus layers, the concentration of adatoms, and the nature of the functionalization and try to emphasize the anisotropy of the material. In contrast to other theoretical research conducted through DFT calculations \cite{Ding2015,Valagiannopoulos2017,MARIAN2022,GOULART2019} and the TB Hamiltonian method \cite{Yuan2015}, our research stands out by its unique emphasis on exploring transport properties. Specifically, we delve into the impact of both covalent and noncovalent functionalization, particularly under conditions of high disorder where the Boltzmann semi-classical approach is not applicable. 

The paper is organized as follows: The atomic structure and the TB Hamiltonian are presented in Sec. \ref{Sec_struc}. In Sect. \ref{sec:NumMethod}, the numerical method for transport calculation is briefly detailed as well as the calculated transport coefficients. Results for resonant scatterers and nonresonant scatterers are presented in Sec. \ref{SecResScat} and \ref{SecNonResScatt}, respectively. Section \ref{SecConclusion} is dedicated to the discussion and conclusion. 
Additional figures are shown in the Supplemental Material \cite{supplementaryMat} (see Appendix 
Sec.\,\ref{Sec_Appendix}).

\section{\label{Sec_struc} Phosphorene structure and effective Hamiltonian}

\begin{figure}
\includegraphics[width=0.48\textwidth]{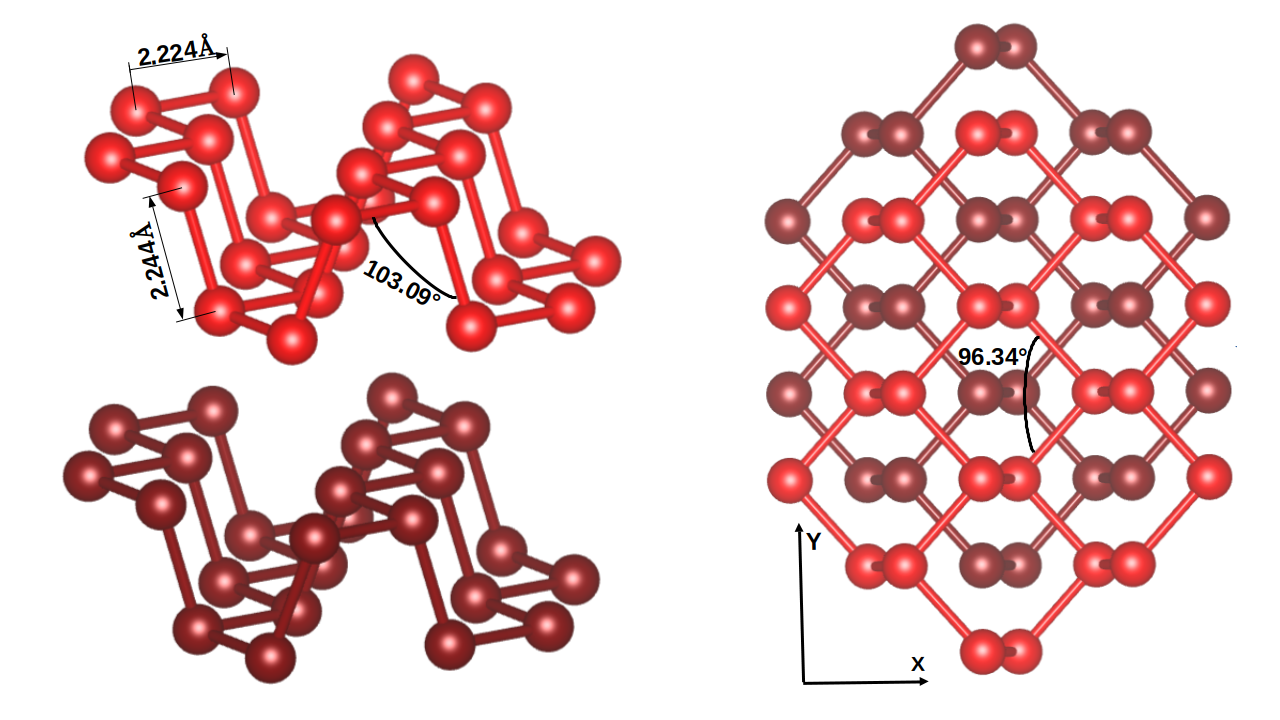}

\caption{Atomic structure of bilayer black phosphorene (BP). The two planes have different colors.
Left: Perspective side view,
Right: Top view.
\label{Fig_structure}}
\end{figure}

In phosphorene, phosphorus atoms are joined by covalent bonds with bond angles of 96.34$^\circ$ and 103.09$^\circ$ \cite{liu2021chemical, antonatos2022black} (Fig.\,\ref{Fig_structure}). 
The intralayer bonding results from the $sp^3$ hybridization of P atoms, giving rise to three bonding orbitals per two atoms augmented by lone pairs associated with each atom. 
This results in strong stability in the crystal structure because these angles are close to perfect tetragonal ones. 
The bonds out of plane have an angle of approximately 45$^\circ$ degrees with respect to the plane and are responsible for the puckered structure of phosphorene, in contrast with graphene and 2D metal chalcogenides. The first two neighbor atoms in the plane are at distances of 2.224\,Å, and the third neighbor out of  plane is at 2.244\,Å. The crystal structure of black phosphorus is orthorhombic with lattice parameters of 4.376\,Å, 3.314\,Å, and 10.478\,Å for $a$, $b$, and $c$, respectively \cite{Takao1981}. 
In phosphorene, the primitive cell contains four atoms at positions $\pm$ $\left(ua, 0, vc\right) $ and  $\pm$ $\left( (1/2 - u)a, b/2, vc\right) $ with $u = 0.08056$ and $v = 0.10168$. The puckered honeycomb lattice formed by covalently bonded phosphorus atoms results in significant in-plane anisotropy, similar to bulk BP, distinguishing it from many other isotropic 2D materials \cite{Carvalho2016,Batmunkh2016}. 
This anisotropy has implications for electronic properties of phosphorene such as energy band structure, conductivity, etc. \cite{he2015exceptional,Qiao2014,liu2016mobility,Haratipour2018}. 
The $x$ axis is the armchair direction, and the $y$ axis is the zigzag direction (Fig.\,\ref{Fig_structure}).

To study the effect of random disorder, one needs to perform numerical calculations in a supercell containing several hundred thousand atoms. That can only be done in the framework of a TB Hamiltonian. However, in contrast to graphene which involves only one hopping parameter between $p_z$ orbitals, the phosphorene cell has four atoms with four orbitals per atom, $s$, $p_x$, $p_y$, and $p_z$, that need to be considered. The exact modeling of the TB Hamiltonian is a challenge because of the mixture of the orbital $s$ and orbitals $p$. However, it has been shown \cite{skachkova2017electronic,li2015structures,Rudenko2014} that the $p_z$ orbital has a large contribution to the valence and conduction band around the band gap, and its role is predominant compared to the roles of the other orbitals $s$, $p_x$, and $p_y$. Therefore, following Rudenko {\it et al.} \cite{Rudenko2014,Rudenko2015}, it is possible to consider only one orbital per atom that looks like a $p_z$ without being exactly the $p_z$ orbital. This approximation is reasonable as long as the energy is fairly close to the band gap.

Indeed, this effective model has been successfully used to describe studies concerning phosphorene nanoribbons \cite{Ezawa_2014,Sisakht2015,vahedi2021edge}, electric \cite{Ezawa_2014,Dolui2015} and magnetic fields \cite{Pereira2015,Yuan2015,Wu20}, field-effect electronic devices \cite{Li2014}, and quantum dots \cite{abdelsalam2018multilayer}. 
We also use the effective TB model which corresponds to one orbital $\vert i \rangle$ per P site \cite{Rudenko2015},

\begin{equation}
{H} =\sum_{i} \epsilon_{i} \vert i \rangle \langle i \vert + \sum_{i\neq j}
t^{\parallel}_{ij} \vert i\rangle\langle j \vert 
+ \sum_{i\neq j} t^{\perp}_{ij} \vert
i \rangle \langle j \vert,
\label{Eq_H}
\end{equation}
where $i$ is the orbital located at $\bf{r}_{i}$ and the sum runs over neighboring $i$, $j$ lattice sites. $\epsilon_{i}$ is the on-site energy of the electron and it takes the same value for all sites because we suppose that all sites are equivalent. $t^{\parallel}_{ij}$ ($ t^{\perp}_{ij}$) is the intralayer (interlayer) hopping parameter between the $i$ and $j$ sites. Those parameters are obtained \cite{Rudenko2015} by the construction of a set of four maximally localized Wannier functions corresponding to $p_z$-like orbitals from DFT calculation with quasiparticle $GW$ approximation \cite{Rudenko2015}. 
We use the TB parameters proposed by Rudenko {\it et al.} in Ref. \cite{Rudenko2015}: ten intralayer and four interlayer hopping parameters over distances between the corresponding interacting sites up to 5.5 Å.

\begin{figure}
\includegraphics[width=0.48\textwidth]{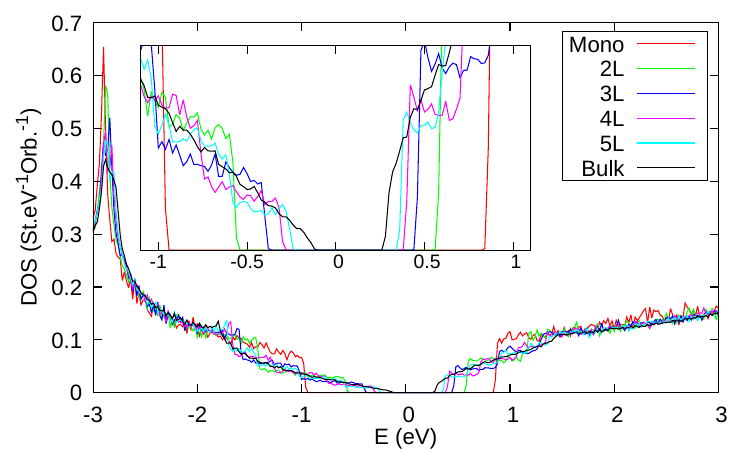}		
\caption{Tight binding (TB) density of states (DOS) of defect-free phosphorene, few-layer BP (up to 5 layers), and bulk BP, calculated using the TB parameters proposed by Ref.\,\cite{Rudenko2015}.}
\label{Fig_DOSpure}
\end{figure}

Figure\,\ref{Fig_DOSpure} shows the total DOS of defect-free multilayers of BP (from monolayer to bulk). One can see that the band gap of the monolayer is about 1.75 eV.
The origin of this gap is the fact that this TB Hamiltonian involved the nearest neighbor atoms, two in-plane hopping terms $t_1=-1.22$ eV, and one out-of-plane hopping term $t_2=3.67$ eV. The distinct sign of $t_2$ in comparison to $t_1$ implies the opening of the gap \cite{Rudenko2014,Rudenko2015,ma2021anomalous}. 
It should be noted that more theoretical works have predicted an energy gap value ranging from 1 eV \cite{Liu2014,Qiao2014,Peng2014} up to 2 eV \cite{Tran2014} depending on the calculation method, and the measured value experimentally is around 2 eV \cite{Gaufres19, Castellanos2014,Wang2015}.

As already reported experimentally and theoretically \cite{cai2014layer,li2017direct}, the band gap decreases with the number of layers, and for the bulk, its value is around 0.3\,eV. 
This band gap evolution is consistent with an increase of the transversal confinement when the number of layers decreases.
It is also interesting to note that as the number of layers decreases, the electron-hole asymmetry becomes less pronounced.
Indeed, for the monolayer, the DOS is almost symmetric with respect to the band gap. 
Whereas, as the number of layers increases, the DOS of the valency band becomes lower than the DOS of the conduction band (see inset of Fig.\,\ref{Fig_DOSpure}).

\section{\label{sec:NumMethod} Numerical method for transport}
The electronic properties of the multilayer phosphorene are calculated using the real-space Kubo-Greenwood formalism computed by recursion and polynomial expansion method \cite{Mayou1988,Mayou95,Roche1997,Roche99,Triozon2002}. This approach efficiently computes quantum transport in large systems for which diagonalization is not numerically feasible. The formalism has been employed to study the effect of vacancies, defects, and disorder on quantum transport in various materials, including monolayer and bilayer graphene \cite{Missaoui2017,Missaoui2018,Trambly13,Lherbier2012,Roche2012,Cresti2013,Namarvar2020}, carbon nanotubes \cite{Latil2004,Ishii2010,Jemai2019}, quasicrystals \cite{Triozon2002}, and perovskites \cite{Lacroix20}.

This real-space method enables the inclusion of local defects directly into the Hamiltonian and allows for a random distribution of defects in the huge supercell. To systematically study the effect of defects on electronic properties, we investigate two types of generic defects:

(i) Resonant defects, {\it i.e.}~defects that create a covalent bond with a P atom of phosphorene. As the system is described by the effective four-band Hamiltonian, with one $p_z$-like orbital per phosphorus atom, we employ the same approach as in the case of graphene to model resonant scatterers \cite{Trambly13, Missaoui2017, Missaoui2018}. An adatom or ad-molecule bound covalently with the P atom is almost equivalent for the electron bands to removing the corresponding orbitals of the bonded P. Thus, a resonant scatter is modeled by a vacant P atom. In our calculations, defects (vacancies) are randomly distributed over the entire supercell of 640000 P atoms per layer.

(ii) Nonresonant defects, {\it i.e.}~noncovalent functionalization, such as van der Waals bonding, that create a small local perturbation of the electronic structure. A common method for dealing with this type of defect is to use Anderson disorder, where the on-site energies $\epsilon_i$ 
of all sites (Eq.\,(\ref{Eq_H}))  are randomly distributed in $[\epsilon_i^0 - W/2, \epsilon_i^0 + W/2]$. This uniform Anderson disorder is one of the simplest models (with only one parameter) that is commonly used for nonresonant defects \cite{Lee85}. Here, $W$ is the magnitude of that disorder, and $\epsilon_i^0$ is the on-site energy of the orbital $i$ without disorder from Ref.\,\cite{Rudenko2015}.

To evaluate the effect of the anisotropy of phosphorene on the conductivity $\sigma$, we compute \cite{Mayou1988,Mayou95,Roche1997,Roche99,Triozon2002} $\Delta X^{2}(E,t)$ and $\Delta Y^{2}(E,t)$, which represent the average square of the quantum spreading along the armchair direction and the zigzag direction, respectively (see Fig.\,\ref{Fig_structure}), after a time $t$ and for states at energy $E$.

In such calculations, all quantum effects, including all multiple-scattering effects, are taken into account to calculate the average square spreading $\Delta X^2$ without inelastic scattering, i.e., at zero temperature. 
At finite temperature $T$, the inelastic scattering caused by the electron-phonon interactions is accounted for using the relaxation time approximation (RTA) \cite{Trambly13}. The conductivity in the $x$-direction ($y$-direction) is given by,
\begin{eqnarray}
\sigma(E_{F},\tau_{i}) &=& e^{2}n(E_{F})D(E_{F},\tau_{i}) , \label{Eq_Sigma}\\
D(E_{F},\tau_{i}) &=& \frac{L_{i}^{2}(E_{F},\tau_{i})}{2\tau_{i}} ,\label{Eq_D_tau}\\
L_{i}^{2}(E_{F},\tau_{i}) &=& \frac{1}{\tau_{i}}\int_{0}^{\infty}\Delta X^{2}(E_F,t)e^{{-t}/{\tau_{i}}}{\rm d}t, \label{Eq_Li}
\end{eqnarray}
where $E_F$ is the Fermi energy, $\tau_i$ is the inelastic scattering time, $n(E)$ is the total DOS, $D$ is the diffusivity along the $x$ axis ($y$ axis), and $L_{i}$ is the inelastic mean free path. $L_{i}(E_F,\tau_{i})$ represents the typical distance of propagation during the time interval $\tau_{i}$ for electrons at energy $E$. In simple terms, $\tau_i$ is the time beyond which the velocity autocorrelation function exponentially approaches zero\,\cite{Trambly13}.

\begin{figure*}
\begin{center}
\includegraphics[width=0.95\textwidth]{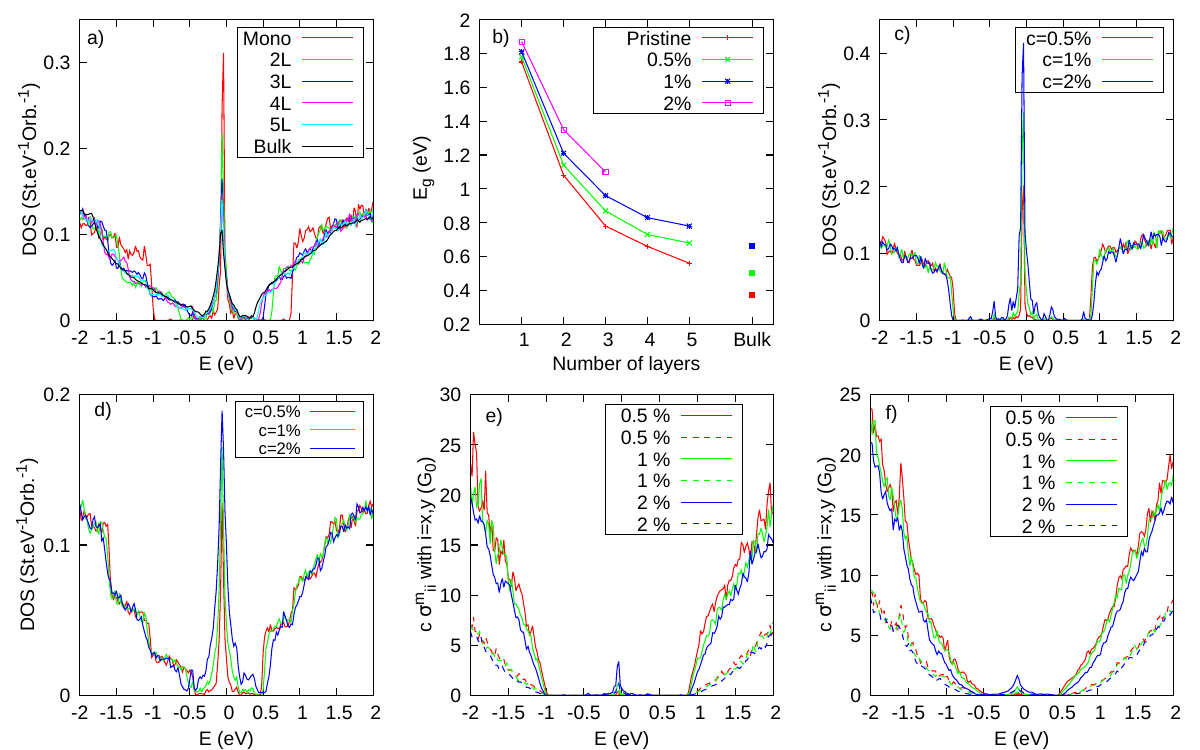}
\caption{\label{Fig_Res_Lac}
With resonant scatterers (vacancies):
(a) DOS for defect concentration $c=1\,\%$ depending on the number of layers for phosphorene,
(b) $E_g$ for different $c$ and for 1--5 layers of phosphorene and bulk,
(c) monolayer DOS for different $c$,
(d) trilayer DOS for different $c$.
(e) (f) Microscopic conductivity (Eq.\,(\ref{Eq_SigmaM})) times defect concentration $c$,
(bold line) $c\,\sigma^{\rm m}_{xx}$ 
and (dashed line) $c\,\sigma^{\rm m}_{yy}$, 
versus $E$, in (e) monolayer and (f) trilayer.
These results show that $c\,\sigma^{\rm m}$ is almost independent of $c$ for energies outside the gap, which proves Boltzmann-type behavior.
$G_0 = 2 e^2/h$.}
\end{center}	
\end{figure*} 

\begin{figure*}
\includegraphics[width=0.95\textwidth]{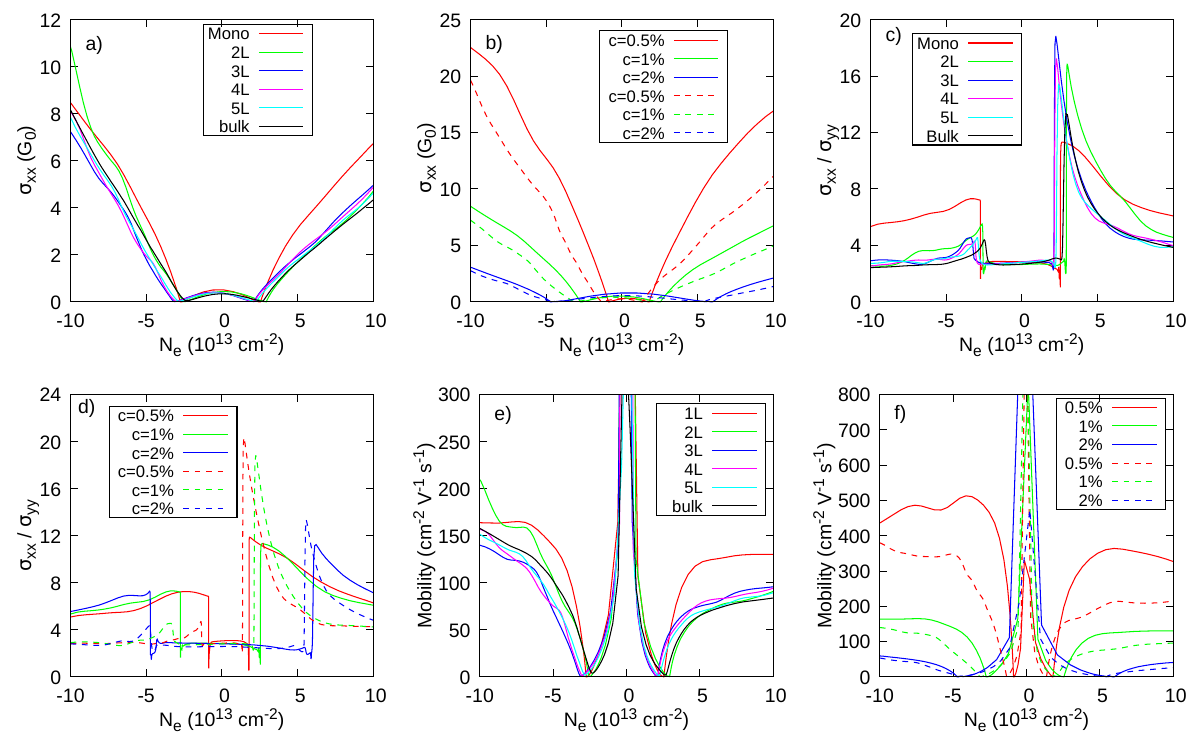}
\caption{\label{Fig_Cond_ft_Ne_lacunes}
\mycomment{
\textcolor{blue}{\st{Microscopic conductivity}}
}
Conductivity at $T=300$\,K (Eq.\,(\ref{Eq_sigma300K})) per layer with resonant scatterers (vacancies): 
(a) $\sigma_{xx}$ versus $N_e$ for 1--5 layers and bulk and for $c=1$\,\%,
(b) $\sigma_{xx}$ for the monolayer (bold line) and trilayer (dashed line) for different defect concentrations $c$,
(c) $\sigma_{xx}/\sigma_{yy}$ versus $N_e$ for $c=1$\,\%,
(d) $\sigma_{xx}/\sigma_{yy}$ versus $N_e$ for the monolayer (bold line) and trilayer (dashed line),
(e) Mobility versus $N_e$ for 1--5 layers and bulk and for $c=1\%$,
(f) Mobility for the monolayer (bold line) and trilayer (dashed line) for different concentrations $c$.
$N_e$ is the density of charge carriers per layer.
$G_0 = 2 e^2/h$. }
\end{figure*}

\begin{figure*}
\includegraphics[width=0.95\textwidth]{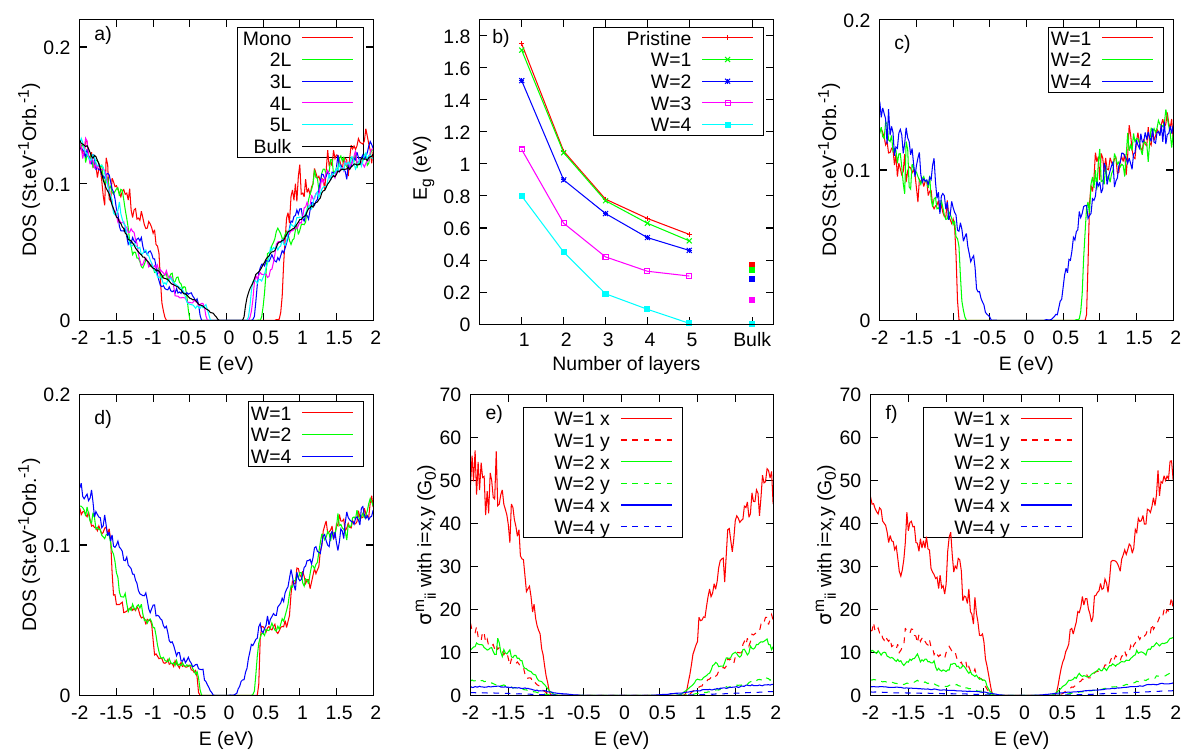}
\caption{\label{Fig_Res_Anderson}
With nonresonant scatterers (Anderson disorder):
(a) Density of states for $W=1$\,eV depending on the number of layers for phosphorene.
(b) $E_g$ for different Anderson disorders and for 1--5 layers of phosphorene and bulk. 
DOS for different $W$\,(eV) values in (c) the monolayer and (d) the trilayer.
(e) (f) Microscopic conductivity (Eq.\,(\ref{Eq_SigmaM})) for different Anderson disorders: (bold line) $\sigma^{\rm m}_{xx}$ and (dashed line) $\sigma^{\rm m}_{yy}$ versus $E$, in (e) the monolayer and (f) trilayer.
$G_0 = 2 e^2/h$.}
\end{figure*} 

\begin{figure*}
\includegraphics[width=0.95\textwidth]{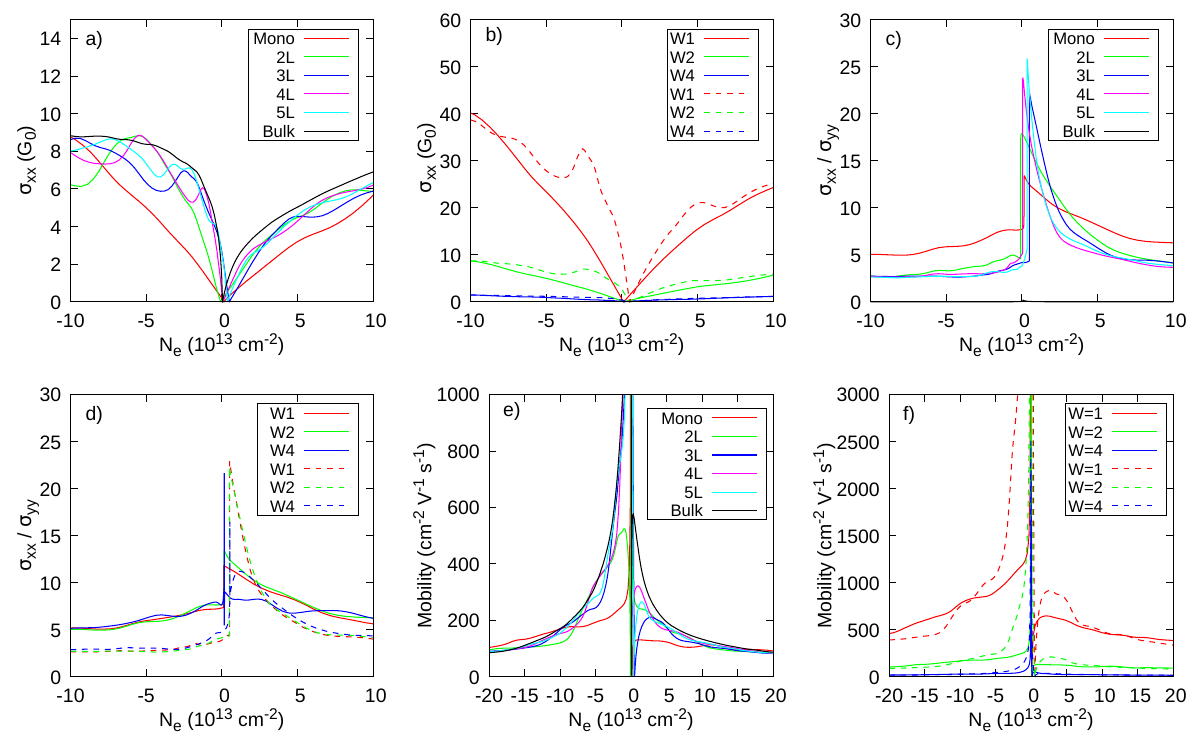}
\caption{\label{Fig_Cond_ft_Ne_Anderson}
Conductivity at $T=300$\,K (Eq.\,(\ref{Eq_sigma300K})) per layer with nonresonant scatterers (Anderson disorder): 
(a) $\sigma_{xx}$ versus $N_e$ for 1--5 layers and bulk, and for Anderson disorder $W=1$\,eV, 
(b) $\sigma_{xx}$ for monolayer and bilayer for different magnitudes $W$\,(eV) of disorders, 
(c) $\sigma_{xx}/\sigma_{yy}$ versus $N_e$ for $W=1$\,eV,
(d) $\sigma_{xx}/\sigma_{yy}$ versus $N_e$ for monolayer and trilayer,
(e) Mobility versus $N_e$ for 1--5 layers and bulk, and for $W=2$\,eV,
(f) Mobility for monolayer (bold line) and trilayer (dashed line) for different Anderson disorders. 
$N_e$ is the density of charge carriers per layer. 
$G_0 = 2 e^2/h$.}
\end{figure*}

As the monolayer and its components are semiconductors, we use Gaussian broadening of the spectrum to compute the DOS, avoiding the tail expansion of Lorentzian broadening \cite{Jouda21}. 
In the present paper, a Gaussian broadening with a standard deviation of $10$\,meV is used. 

The microscopic conductivity $\sigma^{\rm m}(E)$ at energy $E$ is defined as the maximum value of the conductivity, calculated from ${\rm Max}\{D(E,\tau_i)\}_{\tau_i}$:
\begin{equation}
\sigma^{\rm m}(E) = {\rm Max}\{\sigma(E)\}_{\tau_i} .
\label{Eq_SigmaM}
\end{equation}

The microscopic conductivity at energy $E$ (Eq.\,(\ref{Eq_SigmaM})) includes the effects of temperature on the diffusivity of charge carriers for high temperatures or room temperature \cite{Trambly13}, but it does not take into account the mixing of states of various energies $E$ due to the Fermi-Dirac distribution. 
Therefore, we calculate the conductivity at room temperature $\sigma$ as follows:
\begin{equation}
\sigma(\mu_c,T \approx 300\,{\rm K}) = -\int_{-\infty}^{+\infty} \sigma^{\rm m}(E) \frac{\partial f(\mu_c)}{\partial E} {\rm d}E,
\label{Eq_sigma300K}
\end{equation}
where $\mu_c$ is the chemical potential, $\sigma^{\rm m}(E)$ is calculated from Eq.\,\ref{Eq_SigmaM}, and $f$ is the Fermi-Dirac distribution function.

At temperature $T$, the mobility $\mu$ is related to the conductivity $\sigma$ by:
\begin{equation}
\sigma(\mu_c,T) = e \lvert N_e(\mu_c,T) \rvert \mu(\mu_c,T),
\label{Eq_lien_mu_sigma}
\end{equation}
where $N_e$ is the density of charge carriers with respect to the neutral case at $T=0$. Thus, at room temperature ($T \approx 300$\,K):
\begin{equation}
\mu(\mu_c,T) = \frac{-1}{e \lvert N_e \rvert} \int_{-\infty}^{+\infty} \sigma^{\rm m}(E) \frac{\partial f(\mu_c)}{\partial E} {\rm d}E,
\label{Eq_mobilite}
\end{equation}
with
\begin{equation}
N_e(\mu_c,T) = \frac{1}{S}\int_{-\infty}^{+\infty} n(E) f(E,\mu_c,T) {\rm d}E - N_e^0,
\label{Eq_N_e}
\end{equation}
and
\begin{equation}
N_e^0 = \frac{1}{S}\int_{-\infty}^{+\infty} n(E) f(E,\mu_c=E_{F}^{0},T=0) {\rm d}E,
\end{equation}
where $S$ is the surface and $E_{F}^0$ is the Fermi energy of the system without static defects, thus $E_{F}^{0}$ is at the middle of the gap.

Note that in the presence of defect band states in the gap that has a nonzero $\sigma^{\rm m}$, the DOS at $E_F$ can be non-zero, and therefore the mobility is not well defined numerically since the expression (\ref{Eq_mobilite}) may diverges for $N_e$ approaches zero. This is because the mobility is defined for a semiconductor but not for systems with states at $E_F$.

\section{\label{sec:level4} Results}
In this section, we present the results on the effect of vacancies (resonant scatterers) and Anderson disorder (nonresonant scatterers) on the electronic properties of monolayer BP, few-layer BP, and bulk BP.

\subsection{Resonant scatterers}
\label{SecResScat}

Figure\,\ref{Fig_Res_Lac}(a) illustrates the density of states (DOS) for a vacancy concentration of $c=1\%$ (resonant scatterers) relative to the total number of atoms in monolayer, few-layer BP, and bulk BP. Notably, there is a gap decrease as the number of layers increases.
In line with defect-free behavior, the gap decreases as the number of layers increases.

We can clearly distinguish the midgap state peak due to vacancies in the gap (Fig.\,\ref{Fig_Res_Lac}(a)). It is not centered exactly at zero energy because the coupling terms beyond the first neighbors break the electron-hole symmetry of the bipartite Hamiltonian. 
Its width remains almost constant as the number of layers increases, showing that its energy dispersion is mainly due to intralayer couplings between midgap states. 
As the gap decreases with an increasing number of layers, the specific value of $c$ at which the midgap peaks fill the gap depends on the number of layers. 
For the monolayer, at the studied $c$ concentration, the gap is not filled, and we can distinguish secondary peaks 
(see Figs.\,\ref{Fig_Dos_mono_vac1} in the SM \cite{supplementaryMat}). 
On the other hand, the gap is filled for $c>1-2\%$ in the case of 5 layers and the bulk, respectively 
(see Figs.\,\ref{Fig_DOS_5L_lacunes} and \ref{Fig_DOS_5L_Bulk} in the SM \cite{supplementaryMat}).
In Fig.\,\ref{Fig_Res_Lac}(b), the points corresponding to these particular cases are not represented.

Figure\,\ref{Fig_Res_Lac}(b) shows that overall, for few-layer BP (Fig.\,\ref{Fig_Res_Lac}(d) for 3 layers) and bulk configurations, the gap increases as the vacancy concentration rises, except for the monolayer case where the gap remains nearly constant (Fig.\,\ref{Fig_Res_Lac}(c)).
That nonstandard behavior shows repulsion of continuum states by defect states (midgap states). 
From the point of view of a perturbative analysis, it is reasonable to think that this effect will be greater if the energies of the continuum states are closer to those of the midgap states, i.e. for a smaller gap corresponding to a larger number of layers. 
One way of understanding this phenomenon is perhaps to consider the fact that resonant defects distributed over a sub-lattice of a bipartite lattice can create a gap that increases with the concentration of defects \cite{Missaoui2018,Jouda21}. 
A detailed study would be needed to confirm this, but it is beyond the scope of this paper.

Figures\,\ref{Fig_Res_Lac}(e,f) show the microscopic conductivity multiplied by concentration, {\it i.e.}~$c \sigma^{\rm m}$, versus energy for monolayer and trilayer according to the $x$-direction $\sigma^{\rm m}_{xx}$ (bold line) --armchair direction-- and according to the $y$-direction $\sigma^{\rm m}_{yy}$ (dashed line) --zigzag direction--. See also 
Fig.\,\ref{Cond_lacunes_bulk_ft_E} in SM \cite{supplementaryMat}
for bulk BP.  
The armchair conductivity $\sigma^{\rm m}_{xx}$ is greater than the zigzag conductivity $\sigma^{\rm m}_{yy}$ for all cases (monolayer, few-layer BP, and bulk). The proportionality between $\sigma^{\rm m}_{xx}$ and $\sigma^{\rm m}_{yy}$ remains the same for all systems. This ratio is a signature of the structural anisotropy of these materials, which results from an anisotropy of the effective masses of electrons and holes 
\cite{Qiao2014,Carvalho2016,Xia2014,Akhtar2017}.

Moreover, Figs.\,\ref{Fig_Res_Lac}(e,f) show that the conductivity behaves according to the Bloch-Boltzmann approach, and then the conductivity at every $E$ is proportional to $1/c$. 
This result is consistent with experimental measurements of the monolayer BP grafted with PtCl$_2$ groups \cite{sun2019transport}.
It is also important to note that midgap states do not contribute much to conduction. Indeed, even if midgap states fill the gap, there is still a mobility gap. This may not be true for very high defect concentrations, which would allow strong conduction through defect states that would percolate into the structure to form a kind of midgap-state bands\,\cite{Jouda21}.

We now study the effect of vacancies on the conductivity and mobility at room temperature ($T=300$\,K) versus the charge carriers $N_e$ (Eqs.~(\ref{Eq_sigma300K}), (\ref{Eq_mobilite}) and (\ref{Eq_N_e})). The prediction of other works is not similar and depends roughly on the numerical study \cite{Yuan2015}. 
Figure\,\ref{Fig_Cond_ft_Ne_lacunes}(a) shows the armchair conductivity per layer versus $N_e$ for monolayer, few-layer BP, and bulk with a concentration of vacancies of $c=1\%$. The conductivity of holes is larger than that of electrons for high values of charge carriers. 
This difference is a direct consequence of the asymmetry of the DOS around the gap (see Sec.\,\ref{Sec_struc} comments on Fig.\,\ref{Fig_DOSpure}). 
In the cases of multilayer BP, the steeper DOS on the electron side indicates a lower velocity of charge carriers. 
It is therefore clear that the electron-hole asymmetry is much lower in the case of the monolayer for which there is almost no DOS asymmetry.
The zigzag conductivity presents the same behavior 
(see Fig.\,\ref{Condyy_Ne_c1} in SM \cite{supplementaryMat}).
These results are in agreement with the experimental results found by Xia {\it et al.}\,\cite{Xia2014} for a BP thin film.

To compare the effect of the concentration of vacancies, 
Fig.\,\ref{Fig_Cond_ft_Ne_lacunes}(b) presents the armchair conductivity versus $N_e$ for monolayer (bold line) and trilayer (dashed line) with different concentrations. As expected the conductivity decreases when $c$ increases, 
and the electrons/holes asymmetry is seen. 
In Fig.\,\ref{Fig_Cond_ft_Ne_lacunes}(c), the ratio $\sigma_{xx} /\sigma_{yy}$ is shown for monolayer, few-layer BP, and bulk versus $N_e$ for $c=1\%$. $\sigma_{xx}$ is more significant than $\sigma_{yy}$ regardless of the charge carrier nature. 
This anisotropy of BP is well known and has been highlighted theoretically \cite{liu2016mobility,chaves2017} and experimentally \cite{Haratipour2018}.  
This is due to the anisotropy of the bands and therefore of the effective masses in the vicinity of the gap \cite{Qiao2014,Carvalho2016,Xia2014,Akhtar2017}.
It is easy to see that the anisotropy in the behavior of conductivity is more significant for electrons than for holes in all systems.
However, the ratio $\sigma_{xx}/\sigma_{yy}$ is larger for the monolayer than for other systems in the case of holes, whereas in the case of electrons, this ratio becomes smaller compared to that of the multilayers.
This ratio is also independent of the number of vacancies as found previously (Boltzmann transport). 
This is shown in Fig.\,\ref{Fig_Cond_ft_Ne_lacunes}(d) for monolayer and trilayer, and this is valid for all systems. 
Note that\,$\sigma_{xx} / \sigma_{yy}=\mu_{xx} / \mu_{yy}$.

To enable simpler comparisons with experimental results, we plot the mobility at room temperature versus $N_e$ (Fig.\,\ref{Fig_Cond_ft_Ne_lacunes}(e)), according to formulas (\ref{Eq_mobilite}) and (\ref{Eq_N_e}), for monolayer, few layers, and the bulk for $c=1\%$. In the case of the monolayer, the mobility is of the same order of magnitude regardless of the charge carrier type. However, for other systems, it is slightly lower for electrons compared to holes. This result was obtained experimentally for ultrathin black phosphorus with Cu adatoms \cite{koenig2016}. Furthermore, Fig.\,\ref{Fig_Cond_ft_Ne_lacunes}(f) illustrates that the lower the concentration of vacancies, the more pronounced the difference between the mobility of holes and that of electrons. The presence of a midgap state in the gap changes the value of $N_e$ (because defects here are neutral). It is, therefore, possible to compare mobility values, but it is more difficult to compare the values of $N_e$.

\subsection{Nonresonant scatterers}
\label{SecNonResScatt}

The Anderson disorder (nonresonant scatterers, see Sec.\,\ref{sec:NumMethod}) is randomly distributed across all layers of the different phosphorus systems studied. 
Figures\,\ref{Fig_Res_Anderson}(a,b) and 
Fig.\,\ref{DOS_bulk_Anderson} in the SM \cite{supplementaryMat} 
show the DOS for monolayer, multilayers, and bulk BP for different magnitudes $W$ of disorder. It is easy to see that the disorder decreases the gap for all systems (Figs.\,\ref{Fig_Res_Anderson}(c,d)) unlike the case of resonant scatterers.
This is expected for nonresonant defects, which generally create defect states at the band edges\,\cite{Lee85,Jemai2019}. 
For large values of $W$, the gap gradually reduces and becomes zero, for bulk ($E_g \simeq 0.1$\,eV for $W=4$\,eV) and multilayer systems, indicating a semiconductor/semimetal transition 
(see Figs.\,\ref{DOS_5L_Anderson} and \ref{DOS_bulk_Anderson} in the SM \cite{supplementaryMat}).
This result is in good agreement with recent experiments for noncovalent functionalization \cite{kim2015observation,kim2017microscopic,Chen_2020}, which validates our hypothesis that the Anderson disorder can model noncovalent adsorbates.

In the SM \cite{supplementaryMat}, we show the effect of the Anderson disorder only in the top layer of a few-layer BP on the band gap 
(Fig.\,\ref{Dos_5p_surface_Ander}).
This shows that the effect of disorder on the gap is qualitatively similar to the case with Anderson disorder randomly distributed across all layers.

Figures\,\ref{Fig_Res_Anderson}(e,f) present the armchair microscopic conductivity $\sigma^{\rm m}_{xx}$ and zigzag microscopic conductivity $\sigma^{\rm m}_{yy}$, versus energy for monolayer and for 3 layers. As expected due to anisotropy, $\sigma^{\rm m}_{xx}$ is always larger than $\sigma^{\rm m}_{yy}$ for all values of $E$. It is important to emphasize that the states filling the gap through disorder do not contribute significantly to the conduction. Thus, even though the gap depends quite strongly on the intensity of the disorder, the mobility gap is much less dependent on it.

Figure\,\ref{Fig_Cond_ft_Ne_Anderson} presents the effect of Anderson disorder on the conductivity versus charge carriers $N_e$. The behavior of the conductivity is almost the same for the few-layer BP and the bulk with a slight difference for the monolayer, for low values of $N_e$ (Figs.\,\ref{Fig_Cond_ft_Ne_Anderson}(a,b)). 
We can also see that $\sigma_{xx}$ is higher for holes than for electrons. 
The comparison between $\sigma_{xx}$ and $\sigma_{yy}$ is shown in Figs.\,\ref{Fig_Cond_ft_Ne_Anderson}(c,d). Like the case of resonant scatterers (vacancies), the armchair conductivity is larger than the zigzag conductivity, and the ratio $\sigma_{xx} /\sigma_{yy}$ has the same order as the one obtained in the case of resonant scatterers.

In Figs.\,\ref{Fig_Cond_ft_Ne_Anderson}(e,f), the mobility at room temperature, calculated according to the formula (\ref{Eq_mobilite}), is shown. As for resonant scatterers, such a calculation can produce numerical artifacts for very small $N_e$ values. Indeed, when $N_e$ (Eq.\,(\ref{Eq_N_e})) tends to zero, the conductivity (Eq.\,(\ref{Eq_lien_mu_sigma})) must also tend to zero, which leads to a finite mobility $\mu$. But because of Gaussian Broadening in the calculation of $\sigma$ (Sec.\,\ref{sec:NumMethod}), $\sigma$ is overestimated and $\mu$ may be overestimated for $N_e \simeq 0$. However, this numerical artifact disappears, as soon as $N_e$ is not zero. 
Experimentally, it has been shown that for small solvent molecules, which are noncovalent scatterers, and at concentrations on the order of $10^{13}$\,cm$^{-2}$ and $10^{14}$\,cm$^{-2}$, the hole mobility is higher than that of electrons \cite{Wang2019}. 
That is in good agreement with our results for low $N_e$ values.

\section{Discussions and conclusion}
\label{SecConclusion}
Experimentally, phosphorene often contains a large number of local defects, which are not always well characterized in the literature. These static defects can be due to functionalization by adsorbed atoms or molecules, the effect of the substrate or other structural effects. 
We aim to study the electronic structure and quantum transport in the presence of a high concentration of these defects. 
To do this, we considered two extreme types of generic local defects: resonant and nonresonant defects. 
We have used the Rudenko TB model \cite{Rudenko2014,Rudenko2015} to analyze monolayer black phosphorene (BP), few-layer BP, and bulk BP. The TB model used is an effective four-band Hamiltonian, i.e. one effective orbital per P atom, which is a $p_z$-like orbital. 
This makes it easy to simulate resonant adsorbates by vacancies (vacant P atoms) and nonresonant adsorbates by Anderson disorder (on-site energy disorder).

Concerning the electronic structure, the two types of defects have very different effects on the gap. The resonant scatterers create midgap states that expand when the concentration $c$ of defects increases until the gap is filled. In contrast, nonresonant disorder reduces the gap value by filling the gap edges when its magnitude $W$ increases. As in the case of resonant disorder, a high concentration of nonresonant disorder can fill the gap. This extreme case is easier to achieve for a multilayer BP and the bulk BP, which have a lower gap.

Quantum diffusion calculations show that resonant and nonresonant defects act similarly on electrical conductivity. Indeed, even for the high defect concentrations we studied, the gap states of resonant defects and the gap edge states of nonresonant defects contribute very little to conductivity, and so the mobility gap remains close to the gap of the defect-free system. This suggests that electronic transport should be fairly well described by a semi-classical Bloch-Boltzmann-type approach, even when defects strongly modify the electronic structure. Particularly, the conductivity of states outside the gap is inversely proportional to the resonant defect concentration $c$.
Note that for sufficiently small amounts of defects, when the gap is always present, it is possible to calculate the electron mobility, and we obtain values that are comparable to the experimental results.

As expected, the strong anisotropy of BP's atomic structure induces anisotropy in conductivity and mobility. 
This anisotropy is stronger in the case of monolayer BP. 
From a 2-layer BP, it is pretty close to that of the bulk BP. Our calculations show that this anisotropy persists in the presence of the two types of defects we have studied. 
And here again, it is difficult to find very different behavior between resonant defects and nonresonant defects. 
Comparing our calculations with the conductivity or mobility values found experimentally is difficult, as the defect concentration is not known. However, the anisotropy should be found experimentally. In particular, it would be possible to find experimentally that the ratio $\sigma_{xx} /\sigma_{yy}$ versus the charge carrier density $N_e$ exhibits a well-defined peak for small $N_e$.
We also showed that the hole conductivity is greater than that of the electrons, particularly in multilayers and bulk, and this is valid for both types of defects. We can relate this effect to the asymmetry observed in their DOS without defects.

\section*{Acknowledgments}

Numerical calculations were conducted at LPMC, Universit\'e de Tunis El Manar, and {\it Centre de Calculs} (CDC), CY Cergy Paris Universit\'e. We warmly thank Yann Costes and Baptiste Mary from CDC for their computing assistance. GTL and DM were supported by the ANR projects J2D (ANR-15-CE24-0017) and FlatMoi\,(ANR-21-CE30-0029). JJK acknowledges financial support from CY Advanced Studies.

\bibliography{mabiblio}

\newpage


\onecolumngrid

\section{Appendix}


\setcounter{page}{1}

\label{Sec_Appendix}


\begin{center}

{\large \bf Supplemental Material}

\vspace{0.3cm}
{\large for}

\vspace{0.3cm}
{\Large
Electronic structure and conductivity in functionalized multilayer black phosphorene
}

\vspace{0.5cm}
Jouda Jemaa Khabthani, Khouloud Chika, Ghassen Jema\"{i}, Didier Mayou, Guy Trambly de Laissardi\`ere

\vspace{0.5cm}

\end{center}

\vspace{1cm}
\noindent
In this Supplementary Information, 
additional figures are shown to support arguments given in the main text. 

\renewcommand{\thetable}{S\arabic{table}}
\setcounter{table}{0}

\renewcommand{\theequation}{S\arabic{equation}}
\setcounter{equation}{0}

\renewcommand{\thefigure}{S\arabic{figure}}
\setcounter{figure}{0}

\renewcommand{\thesection}{S\arabic{section}}
\setcounter{section}{0}

\begin{figure}[h]
\begin{center}
\includegraphics[width=0.7\textwidth]{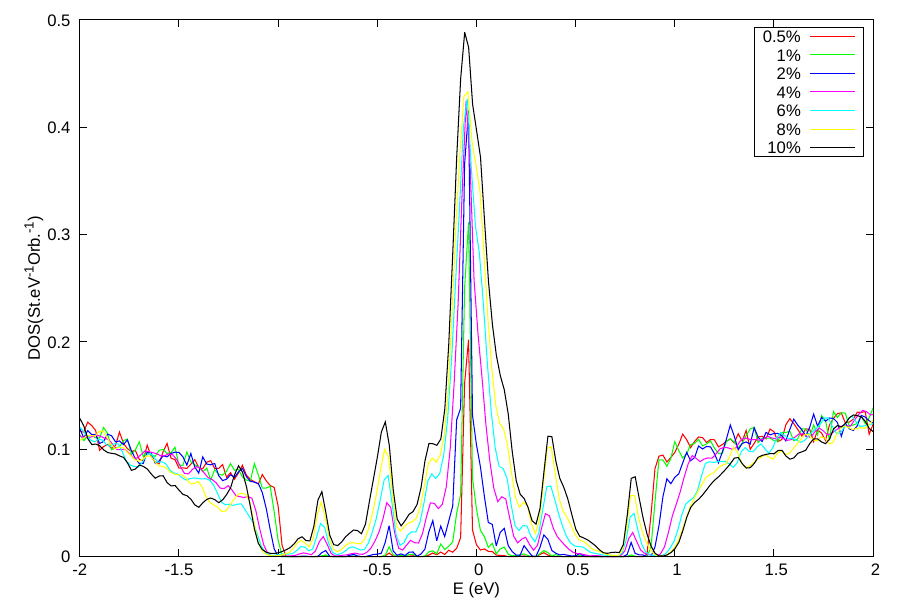}
\caption{Total density of states (DOS) of monolayer BP with different concentrations $c$ (\%) of vacancies. 
\label{Fig_Dos_mono_vac1} }
\end{center}
\end{figure}

\begin{figure}[h]
\begin{center}
\includegraphics[width=0.7\textwidth]{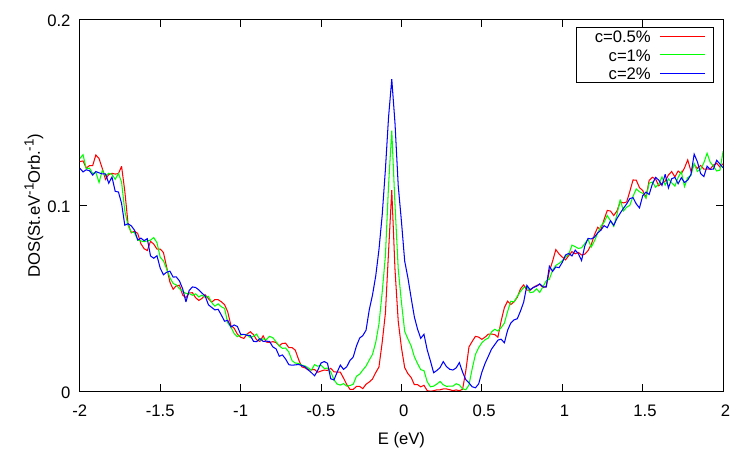}
\caption{Total density of states (DOS) of 5 layers BP  for different concentrations $c$ (\%) of vacancy.}
\label{Fig_DOS_5L_lacunes}
\end{center}
\end{figure}

\begin{figure}[h]
\begin{center}
\includegraphics[width=0.7\textwidth]{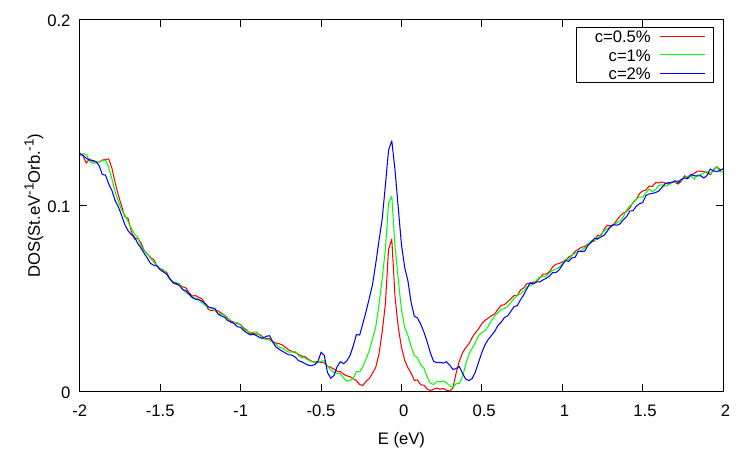}
\caption{Total density of states (DOS) of Bulk BP for different concentrations $c$ (\%) of vacancy.}
\label{Fig_DOS_5L_Bulk}
\end{center}
\end{figure}

\begin{figure}[h]
\begin{center}
\includegraphics[width=0.7\textwidth]{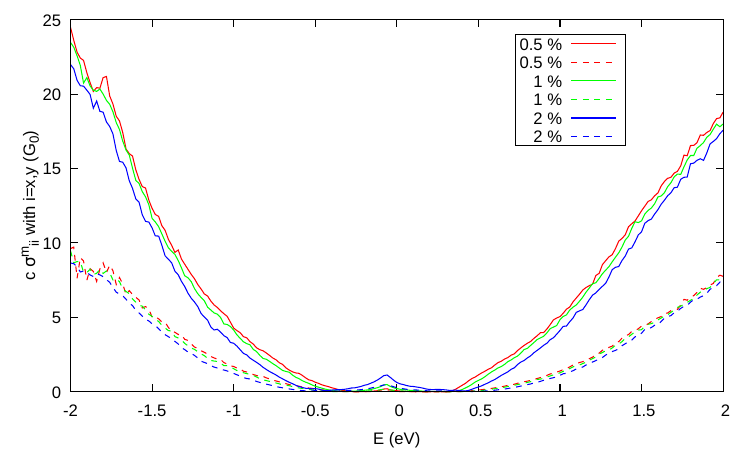}
\caption{Microscopic conductivity (per layer) times $c$ along $x$-, $y$- directions for bulk PB with concentration $c$ of vacancies: (bold line)  $c \,\sigma^{\rm m}_{xx}$ 
and (dashed line) $c \,\sigma^{\rm m}_{yy}$. 
$G_0 = 2 e^2/h$.
}
\label{Cond_lacunes_bulk_ft_E}
\end{center}
\end{figure}

\begin{figure}[h]
\begin{center}
\includegraphics[width=0.7\textwidth]{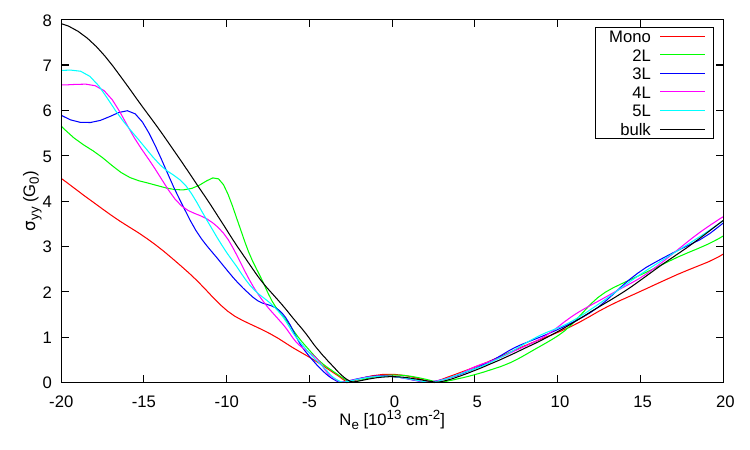}
\caption{Conductivity at $T=300$\,K 
(Eq.\,\ref{Eq_sigma300K})  per layer $\sigma^{\rm m}_{yy}$  of mono, multi and bulk BP with $c=1$\,\% vacancies along $y$-direction versus $N_e$.
$G_0 = 2 e^2/h$.
}
\label{Condyy_Ne_c1}
\end{center}
\end{figure}

\begin{figure}[h]
\begin{center}
\includegraphics[width=0.7\textwidth]{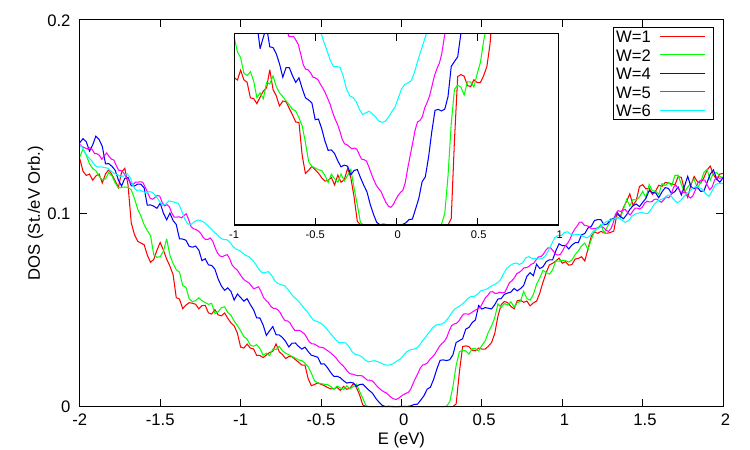}
\caption{ Total density of states (DOS) for 5 layers BP with Anderson disorder of magnitude $W$ (eV).}
\label{DOS_5L_Anderson}
\end{center}
\end{figure}

\begin{figure}[h]
\begin{center}
\includegraphics[width=0.7\textwidth]{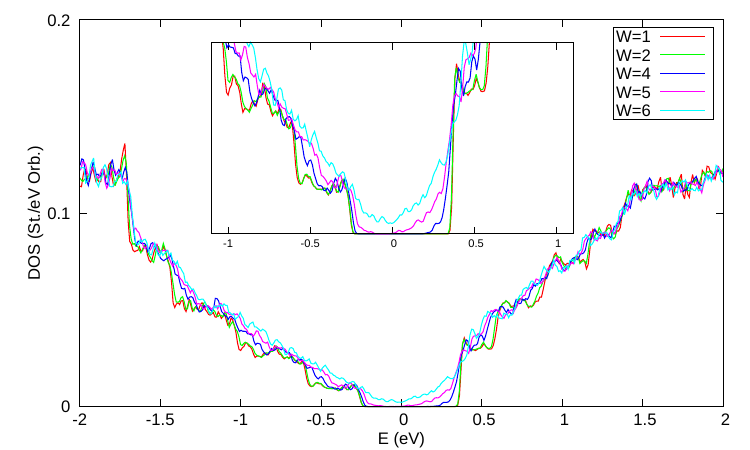}
\caption{Total density of states (DOS) of 5 layers with Anderson disorder in its top-layer of magnitude $W$ (eV).  }
\label{Dos_5p_surface_Ander}
\end{center}
\end{figure}

\begin{figure}[h]
\begin{center}
\includegraphics[width=0.7\textwidth]{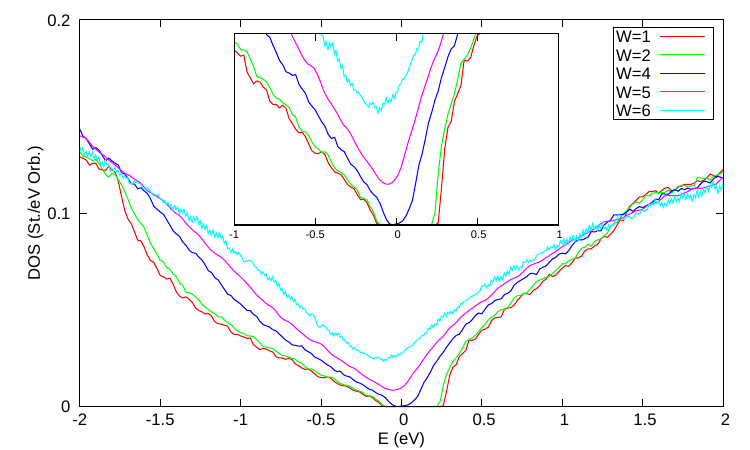}
\caption{Total density of states (DOS) of Bulk BP with different magnitude $W$ (eV) of Anderson disorder.}
\label{DOS_bulk_Anderson}
\end{center}
\end{figure}

\end{document}